\documentstyle[12pt]{article}
\begin{document}

\begin{center}
{\large{\bf A new solution family of the Jacobi equations: \\ characterization, invariants and global Darboux analysis}} \\ 
\mbox{} \\
{\bf Benito Hern\'{a}ndez--Bermejo$^{\; \mbox{\footnotesize {\rm a)}}}$}
\end{center}
{\em Departamento de Matem\'{a}tica Aplicada. Universidad Rey Juan Carlos. } \\
{\em Edificio Departamental II. Calle Tulip\'{a}n S/N. 28933--M\'{o}stoles--Madrid. Spain.}

\mbox{}

\begin{center} 
{\bf Abstract}
\end{center}

\noindent A new family of skew-symmetric solutions of the Jacobi partial differential equations for finite-dimensional Poisson systems is characterized and analyzed. Such family has some remarkable properties. In first place, it is defined for arbitrary values of the dimension and the rank. Secondly, it is described in terms of arbitrary differentiable functions, namely it is not limited to a given degree of nonlinearity. Additionally, it is possible to determine explicitly the fundamental properties of those solutions, such as their Casimir invariants and the algorithm for the reduction to the Darboux canonical form, which have been reported only for a very limited sample of finite-dimensional Poisson structures. Moreover, such analysis is carried out globally in phase space, thus improving the usual local scope of Darboux theorem.  

\mbox{}



\noindent {\bf Running Title:} New Poisson structures.

\mbox{}

\vfill

\mbox{}

\footnoterule
\noindent $^{\mbox{\footnotesize {\rm a)}}}$ E-mail: benito.hernandez@urjc.es 

\pagebreak
\begin{flushleft}
{\bf I. INTRODUCTION}
\end{flushleft}

Finite-dimensional Poisson structures$^{1,2\:}$ are ubiquitous in most domains of mathematical physics, such as fluid dynamics,$^{3\:}$ plasma physics,$^{4\:}$ field theory,$^{5\:}$ continuous media,$^{6\:}$ etc. Finite-dimensional Poisson systems are relevant in the study of very different kinds of nonlinear problems, including population dynamics,$^{7-12\:}$ mechanics,$^{13-16\:}$ electromagnetism,$^{17\:}$ optics,$^{18\:}$ or plasma physics,$^{19\:}$ to cite a sample. The association of a finite-dimensional Poisson structure to a differential system (which is still an open problem$^{16,20-22\:}$) is not only mathematically appealing, but also very useful through the use of a plethora of specialized techniques which include the development of perturbative solutions,$^{17\:}$ numerical algorithms,$^{23\:}$ nonlinear stability analysis by means of either the energy-Casimir$^{20,24\:}$ or the 
energy-momentum$^{25\:}$ methods, characterization of invariants,$^{26\:}$ reductions,$^{2,27\:}$ analysis of integrability properties,$^{28\:}$ establishment of variational principles,$^{29\:}$ study of bifurcation properties and chaotic behavior,$^{18,30\:}$ etc. 

In terms of a system of local coordinates on an $n$-dimensional manifold, Poisson systems of finite dimension have the form:
\begin{equation}
    \label{jmp4nham}
    \dot{x}_i = \sum_{j=1}^n J_{ij} \partial _j H \; , \;\:\; i = 1, \ldots , n
\end{equation} 
Here and in what follows $ \partial_j \equiv \partial / \partial x_j$. The $C^1$ real-valued function $H(x)$ in (\ref{jmp4nham}) is a constant of motion of the system playing the role of Hamiltonian. The $J_{ij}(x)$, called structure functions, are also $C^1$ and real-valued and constitute the entries of an $n \times n$ structure matrix ${\cal J}$. The $J_{ij}(x)$ are 
characterized by two properties. The first one is that they are skew-symmetric:
\begin{equation}
\label{jmp4sksym}
	J_{ij}=-J_{ji} 
\end{equation}
And secondly, they are solutions of the Jacobi partial differential equations (PDEs in what is to follow):
\begin{equation}
     \label{jmp4jac}
     \sum_{l=1}^n ( J_{il} \partial_l J_{jk} + 	J_{jl} \partial_l J_{ki} 
	+ J_{kl} \partial_l J_{ij} ) = 0 
\end{equation}
In equations (\ref{jmp4sksym}) and (\ref{jmp4jac}) indices $i,j,k$ run from $1$ to $n$. 

There are different reasons justifying the importance of the Poisson representation. One is that it provides a wide generalization of classical Hamiltonian systems, allowing not only for odd-dimensional vector fields, but also because a structure matrix verifying (\ref{jmp4sksym}-\ref{jmp4jac}) admits a great diversity of forms apart from the classical constant symplectic matrix. Actually, Poisson systems are a generalization of the classical Hamiltonian systems on which a noncanonical bracket is defined, namely:
\begin{equation}
\label{jmp4cp}
	\{ f(x),g(x) \} = \sum _{i,j=1}^n \frac{\partial f(x)}{\partial x_i} J_{ij}(x)
	\frac{\partial g(x)}{\partial x_j}
\end{equation}
for every pair of differentiable functions $f(x)$ and $g(x)$. The possible rank degeneracy of the structure matrix ${\cal J}$ implies that a certain class of first integrals ($C(x)$ in what follows) termed Casimir invariants exist. There is no analog in the framework of classical Hamiltonian systems for such constants of motion, which have the property of commuting in the sense of (\ref{jmp4cp}) with all differentiable functions. It can be seen that this implies that Casimir invariants are the solution set of the system of coupled PDEs ${\cal J} \cdot \nabla C =0$. The determination of Casimir invariants and their use in order to carry out a reduction (local, in principle) is the cornerstone of the (at least local) equivalence between Poisson systems and classical Hamiltonian systems, as stated by Darboux Theorem$^{1,2}$, which demonstrates that if an $n$-dimensional Poisson manifold has constant rank of value $r=2s$ everywhere, then at each point of the manifold there exist local coordinates $(p_1, \ldots ,p_{s},q_1, \ldots , q_{s},z_1, \ldots , z_{n-r})$ in terms of which the equations of motion become:
\[
	\left\{ \begin{array}{lcl}
	\displaystyle{\dot{q}_i = \frac{\partial H}{\partial p_i} \;\: , \;\:\:
	\dot{p}_i = - \frac{\partial H}{\partial q_i}} & , & i=1, \ldots ,s  \\
	\mbox{} & \mbox{} & \mbox{} \\
	\dot{z}_j = 0 & , & j=1 , \ldots , n-r
	\end{array} \right.
\]
This justifies that Poisson systems can be regarded, to a large extent, as a rightful generalization of classical Hamiltonian systems. This connection is an additional and important advantage of Poisson systems, as far as it accounts for the potential transfer of results and techniques from classical Hamiltonian theory once a given system has been recognized as a Poisson one and the Darboux canonical form has been constructed, specially if this can be done globally. As indicated, the problem of recasting a vector field not written in the form 
(\ref{jmp4nham}) in terms of a finite-dimensional Poisson system is also an open issue of fundamental importance in this context to which important efforts have been devoted in a variety of approaches,$^{7-22\:}$ all of which obviously require the use of solutions of 
(\ref{jmp4sksym}-\ref{jmp4jac}). This, together with the intrinsic mathematical interest of the problem, explains also the attention deserved in the literature by the obtainment and classification of skew-symmetric solutions of the Jacobi equations.$^{7-21,31-34\:}$ 

Due to the reason that equations (\ref{jmp4jac}) constitute a set of coupled nonlinear PDEs, the characterization of solutions of (\ref{jmp4sksym}-\ref{jmp4jac}) has proceeded by means of either suitable {\em ansatzs\/}$^{7-11,32,33\:}$ or through a diversity of other 
approaches.$^{12-16,20-22,31,34\:}$ In particular, there is a clear lack of knowledge of solutions verifying the following six properties: (i) to have arbitrary dimension $n$; (ii) for every $n$, to allow arbitrary (even) values of the rank; (iii) to be defined in terms of arbitrary differentiable functions, namely functions of arbitrary nonlinearity; (iv) a complete set of independent Casimir invariants can be determined; (v) it is also possible to construct the Darboux canonical form; (vi) items iv and v can be carried out globally in phase space. In this work, a new family of skew-symmetric solutions of the Jacobi equations is characterized and analyzed. Such family shall be termed multiseparable in what follows, due to its functional form, reminiscent of a multiple separation of variables. This family presents the remarkable feature of complying to all the conditions (i)-(vi) just enumerated. According to such criterion, it can be regarded as a significant contribution in this context. In particular, it is worth noting also that previously known types of Poisson structures appearing in a diversity of physical situations and systems can be seen to be obtainable as particular cases of the new multiseparable family of solutions, as it will be illustrated in the examples section. Moreover, the constructive and global Darboux analysis to be presented constitutes an improvement of the usual scope of Darboux theorem, which does only guarantee in principle a local reduction,$^{1,2\:}$ as mentioned. In addition, the achievement of such reduction is relevant as far as the explicit determination of the Darboux coordinates is often a complicated task, only known for a limited sample of finite-dimensional Poisson 
structures.$^{2,8,27,31\:}$ 

The article has the following structure. In Section II the new solutions are characterized. 
Their symplectic structure and the reduction to the Darboux canonical form are determined in Section III. Examples are provided in Section IV. Section V concludes with some final remarks. 

\mbox{}

\begin{flushleft}
{\bf II. CHARACTERIZATION OF THE SOLUTION FAMILY}
\end{flushleft}

We begin with a preliminary definition: 

\mbox{}

{\em Definition 2.1: \/} Let $A =(a_{ij})$ and $B =(b_{ij})$ be two $n \times n$ real and regular matrices ($n \geq 2$) such that $A = B^{-1}$. Let also $B_i \equiv (b_{i1}, \ldots ,b_{in})$ denote the $i$-th row of $B$, for $i=1, \ldots ,n$. In addition, let $\Omega \subset I \!\! R^n$ be a domain in which a system of local coordinates $x=(x_1, \ldots ,x_n)$ is defined. If $r$ is an even integer, $2 \leq r \leq n$, we shall denote by $\Omega_i^* \subset I \!\! R$ the subsets $\Omega_i^* \equiv \{ B_i \cdot x : x \in \Omega \}$, for $i=1, \ldots ,r$. Let also $\psi_i(x) : \Omega \rightarrow I \!\! R$, with $i=1, \ldots ,r$, denote $r$ functions which are $C^1(\Omega)$ and do not vanish at any point of $\Omega$, and such that they can be expressed in the form $\psi_i(x) = \varphi_i(B_i \cdot x)$, where every function $\varphi_i(y_i) : \Omega ^*_i \rightarrow I \!\! R$ is $C^1(\Omega^*_i)$ and does not vanish in any point of $\Omega^*_i$. Finally let 
\begin{equation}
\label{jmp4car}
	\Lambda_{ij}^{kl} \equiv \left| \begin{array}{cc} 
	a_{ik} & a_{il} \\ a_{jk} & a_{jl} \end{array} \right| = a_{ik}a_{jl} - a_{il}a_{jk}
	\:\; , \;\:\;\: i,j,k,l = 1 , \ldots , n,
\end{equation}
where the vertical lines denote hereafter a matrix determinant. Then a $n \times n$ matrix ${\cal J}(x) \equiv (J_{ij}(x))$ defined in $\Omega$ is termed multiseparable if it has the form:
\begin{equation}
\label{jmp4mss}
	J_{ij}(x)\equiv J_{ij}^{[r]}(x) = \sum_{k=1}^{r/2} \Lambda_{ij}^{2k-1,2k} 
	\psi_{2k-1}(x) \psi_{2k}(x) \:\; , \;\:\;\: i,j = 1 , \ldots , n
\end{equation}
Moreover, for every $n \geq 2$, multiseparable matrices will be also defined in $\Omega$ for the additional even value $r=0$ as $J_{ij}^{[0]}(x)=0$ for $i,j=1, \ldots ,n$ and for every $x \in \Omega$.

\mbox{}

This definition provides the basis for the following result:

\mbox{}

{\bf Theorem 2.2: \/} Let $n \geq 2$ be an integer, and let $\Omega \subset I \!\! R^n$ be a domain in which a multiseparable matrix ${\cal J}$ is defined. Then ${\cal J}$ is a structure matrix globally defined in $\Omega$. 

\mbox{}

{\em Proof: \/} Since the case $r=0$ is clear, we shall focus on the case $r \geq 2$. 
Skew-symmetry of ${\cal J}$ is a consequence of the fact that $\Lambda _{ij}^{2k-1,2k}=- \Lambda _{ji}^{2k-1,2k}$ for $i,j = 1, \ldots ,n$ and for $k=1, \ldots ,r/2$ in 
(\ref{jmp4car}-\ref{jmp4mss}). Let us now turn to the Jacobi identities (\ref{jmp4jac}). Substitution of (\ref{jmp4mss}) into (\ref{jmp4jac}) produces after some rearrangements:
\[
     \sum_{l=1}^n ( J_{il} \partial_l J_{jk} + 	J_{jl} \partial_l J_{ki} 
	+ J_{kl} \partial_l J_{ij} ) = 
\]
\[
	\sum_{p,q=1}^{r/2} \varphi_{2p-1}\varphi_{2p} 
	\left\{ \varphi_{2q-1}^{\prime} \varphi_{2q} \sum_{l=1}^n b_{2q-1,l}
	\left( \Lambda_{il}^{2p-1,2p}\Lambda_{jk}^{2q-1,2q} + 
	\Lambda_{jl}^{2p-1,2p}\Lambda_{ki}^{2q-1,2q}+
	\Lambda_{kl}^{2p-1,2p}\Lambda_{ij}^{2q-1,2q} \right) \right.
\]
\[
	+ \varphi_{2q-1} \varphi_{2q}^{\prime} \left. \sum_{l=1}^n b_{2q,l}
	\left( \Lambda_{il}^{2p-1,2p}\Lambda_{jk}^{2q-1,2q} + 
	\Lambda_{jl}^{2p-1,2p}\Lambda_{ki}^{2q-1,2q}+
	\Lambda_{kl}^{2p-1,2p}\Lambda_{ij}^{2q-1,2q} \right) \right\} \equiv
\]
\begin{equation}
\label{jmp4t12}
	\sum_{p,q=1}^{r/2} \varphi_{2p-1}\varphi_{2p} 
	\left\{ \varphi_{2q-1}^{\prime} \varphi_{2q} T_1 + 
	\varphi_{2q-1} \varphi_{2q}^{\prime} T_2 \right\}
\end{equation}
where $T_1$ and $T_2$ are terms to be examined separately. Let us first look at $T_1$. Using the definition of the constants $\Lambda_{ij}^{kl}$ given in (\ref{jmp4car}), after some algebra it is found that:
\[
T_1 = \left| \begin{array}{cc} a_{i,2p-1} & a_{i,2p} \\ \delta_{2q-1,2p-1} & \delta_{2q-1,2p} \end{array} \right| \cdot 
\left| \begin{array}{cc} a_{j,2q-1} & a_{j,2q} \\ a_{k,2q-1} & a_{k,2q} \end{array} \right| + 
\]
\[
\left| \begin{array}{cc} a_{j,2p-1} & a_{j,2p} \\ \delta_{2q-1,2p-1} & \delta_{2q-1,2p} 
\end{array} \right| \cdot 
\left| \begin{array}{cc} a_{k,2q-1} & a_{k,2q} \\ a_{i,2q-1} & a_{i,2q} \end{array} \right| + \left| \begin{array}{cc} a_{k,2p-1} & a_{k,2p} \\ \delta_{2q-1,2p-1} & \delta_{2q-1,2p} 
\end{array} \right| \cdot 
\left| \begin{array}{cc} a_{i,2q-1} & a_{i,2q} \\ a_{j,2q-1} & a_{j,2q} \end{array} \right| 
\]
where the symbol $\delta_{ij}$ stands for Kronecker's delta. Notice that in $T_1$ it is always $\delta_{2q-1,2p}=0$ since $p$ and $q$ are integers. Now consider two complementary cases for $T_1$:
\begin{description}
\item[{\em Case 1.1.}] Assume $p=q$ in $T_1$. Then $\delta_{2q-1,2p-1}=1$ and $T_1$ becomes:
\[
T_1= a_{i,2p}(a_{j,2p}a_{k,2p-1}-a_{j,2p-1}a_{k,2p})+a_{j,2p}(a_{k,2p}a_{i,2p-1}-
\]
\[
a_{i,2p}a_{k,2p-1})+a_{k,2p}(a_{i,2p}a_{j,2p-1}-a_{i,2p-1}a_{j,2p})=0
\]
\item[{\em Case 1.2.}] Let $p \neq q$ in $T_1$. Now $\delta_{2q-1,2p-1}=0$ and $T_1$ vanishes straightforwardly.
\end{description}
Consequently it is $T_1=0$ in all cases. Similarly, let us now examine $T_2$. Following an analogous procedure it can be found that:
\[
T_2 = \left| \begin{array}{cc} a_{i,2p-1} & a_{i,2p} \\ \delta_{2q,2p-1} & \delta_{2q,2p} 
\end{array} \right| \cdot 
\left| \begin{array}{cc} a_{j,2q-1} & a_{j,2q} \\ a_{k,2q-1} & a_{k,2q} \end{array} \right| + 
\]
\[
\left| \begin{array}{cc} a_{j,2p-1} & a_{j,2p} \\ \delta_{2q,2p-1} & \delta_{2q,2p} 
\end{array} \right| \cdot 
\left| \begin{array}{cc} a_{k,2q-1} & a_{k,2q} \\ a_{i,2q-1} & a_{i,2q} \end{array} \right| + \left| \begin{array}{cc} a_{k,2p-1} & a_{k,2p} \\ \delta_{2q,2p-1} & \delta_{2q,2p} 
\end{array} \right| \cdot 
\left| \begin{array}{cc} a_{i,2q-1} & a_{i,2q} \\ a_{j,2q-1} & a_{j,2q} \end{array} \right| 
\]
As before, note that $\delta_{2q,2p-1}=0$ in $T_2$ since $p$ and $q$ are integers. Two complementary cases appear now for $T_2$:
\begin{description}
\item[{\em Case 2.1.}] It is $p=q$ in $T_2$. Thus $\delta_{2q,2p}=1$ and $T_2$ reduces to:
\[
T_2= a_{i,2p-1}(a_{j,2p-1}a_{k,2p}-a_{j,2p}a_{k,2p-1})+a_{j,2p-1}(a_{i,2p}a_{k,2p-1}-
\]
\[
a_{i,2p-1}a_{k,2p})+a_{k,2p-1}(a_{i,2p-1}a_{j,2p}-a_{i,2p}a_{j,2p-1})=0
\]
\item[{\em Case 2.2.}] Assume $p \neq q$ in $T_2$. Then $\delta_{2q,2p}=0$ and it is immediate that $T_2$ vanishes.
\end{description}
Therefore we also have $T_2=0$ in all cases. Together with the previous result $T_1=0$, this implies in (\ref{jmp4t12}) that multiseparable matrices verify the Jacobi equations 
(\ref{jmp4jac}) for $r \geq 2$. This completes the proof of Theorem 2.2. \hfill Q.E.D.

\mbox{}

One of the most significant features of the multiseparable family of Poisson structures is that it can be explicitly and globally analyzed both for the determination of its Casimir invariants and for the construction of the Darboux canonical form. The development of such issues is the purpose of the next section. 

\mbox{}

\begin{flushleft}
{\bf III. SYMPLECTIC STRUCTURE AND DARBOUX CANONICAL FORM}
\end{flushleft}

In what follows, a theorem summarizing the main features of the multiseparable solutions is provided. The proof of such theorem is constructive:

\mbox{}

{\bf Theorem 3.1: \/} For every $n$-dimensional ($n \geq 2$) Poisson system (\ref{jmp4nham}) 
defined in a domain $\Omega \subset I \!\! R^n$ and such that ${\cal J} \equiv (J_{ij}^{[r]})$ is a multiseparable structure matrix, we have that: 
\begin{description}
\item[{\rm (a)}] Rank(${\cal J}$)$=r$ everywhere in $\Omega$.
\item[{\rm (b)}] The functions
\begin{equation}
\label{jmp4cas}
	C_i(x) = \sum _{j=1}^n b_{ij}x_j \:\; , \:\;\:\; i=r+1, \ldots ,n
\end{equation}
constitute a complete set of functionally independent Casimir invariants of ${\cal J}$ in $\Omega$.
\item[{\rm (c)}] It is possible to perform globally in $\Omega$ the reduction of system 
(\ref{jmp4nham}) to the Darboux canonical form by means of a transformation which is a diffeomorphism in $\Omega$.
\end{description}

\mbox{}

{\em Proof: \/} The proof of the theorem begins with an auxiliary result:

\mbox{}

{\em Lemma 3.2: \/} If ${\cal J} \equiv (J_{ij}^{[r]})$ is a multiseparable structure matrix defined in the domain $\Omega \subset I \!\! R^n$, then functions (\ref{jmp4cas}) form a set of functionally independent Casimir invariants for ${\cal J}$ in $\Omega$.

\mbox{}

{\em Proof of Lemma 3.2: \/} Functional independence can be seen by direct evaluation of the Jacobian matrix of functions (\ref{jmp4cas}):
\begin{equation}
\label{jmp4md}
	\frac{\partial(C_{r+1}(x), \ldots, C_n(x))}{\partial (x_1, \ldots ,x_n)} = 
	\left( \begin{array}{ccc} b_{r+1,1} & \ldots & b_{r+1,n} \\ 
	\vdots & \mbox{} & \vdots \\ b_{n,1} & \ldots & b_{n,n}
	\end{array} \right)
\end{equation}
Thus the Jacobian (\ref{jmp4md}) has constant rank (equal to $n-r$) in $I \! \! R^n$ as a consequence that matrix $B$ is invertible, and accordingly functions (\ref{jmp4cas}) are functionally independent in $\Omega$. In addition, let us demonstrate that such functions are Casimir invariants. If $r=0$ the result is direct. For $r \geq 2$, we evaluate the $i$-th component of the matrix product ${\cal J} \cdot \nabla C_p$ for every $p=r+1, \ldots ,n$:
\begin{equation}
\label{jmp4ker}
	({\cal J} \cdot \nabla C_p)_i = \sum_{j=1}^n J_{ij} \partial _j C_p = 
	\sum_{k=1}^{r/2} \varphi_{2k-1} \varphi_{2k} \sum_{j=1}^n b_{pj} \Lambda _{ij}^{2k-1,2k}
\end{equation}
After some algebra, (\ref{jmp4ker}) amounts to:
\begin{equation}
\label{jmp4ker2}
	({\cal J} \cdot \nabla C_p)_i = \sum_{k=1}^{r/2} \varphi_{2k-1} \varphi_{2k} 
	\left| \begin{array}{cc} a_{i,2k-1} & a_{i,2k} \\ \delta_{p,2k-1} & \delta_{p,2k} 
	\end{array} \right|
\end{equation}
But note that $p=r+1, \ldots ,n$, while $1 \leq k \leq (r/2)$. This implies that in all cases it is $\delta_{p,2k-1}=\delta_{p,2k}=0$, and the expression in (\ref{jmp4ker2}) vanishes. Consequently, it is ${\cal J} \cdot \nabla C_p=0$ for all $p=r+1, \ldots ,n$ and the proof of Lemma 3.2 is complete. \hfill Q.E.D.

\mbox{}

A direct outcome of Lemma 3.2 is that Rank(${\cal J}$)$\leq r$ everywhere in $\Omega$. Let us now demonstrate that, in fact, $r$ is the actual value of the rank:

\mbox{}

{\em Lemma 3.3: \/} If ${\cal J} \equiv (J_{ij}^{[r]})$ is a multiseparable structure matrix defined in the domain $\Omega \subset I \!\! R^n$, then Rank(${\cal J}$)$= r$ everywhere in $\Omega$.

\mbox{}

{\em Proof of Lemma 3.3: \/} According to Definition 2.1, the result is verified if $r=0$. For $r \geq 2$, in order to prove this lemma recall first that under a differentiable change of variables $y = y(x)$, every structure matrix ${\cal J}(x)$ is transformed into a new structure matrix ${\cal J}^*(y)$ according to the rule:
\begin{equation}
\label{jmp4jdiff}
      J^*_{ij}(y) = \sum_{k,l=1}^n \frac{\partial y_i}{\partial x_k} J_{kl}(x) 
	\frac{\partial y_j}{\partial x_l} \:\; , \;\:\;\: i,j = 1 , \ldots , n
\end{equation}
In our case, we shall perform the following change of variables:
\begin{equation}
\label{jmp4cl}
	y_i= \sum_{j=1}^n b_{ij} x_j \:\; , \;\:\;\: i = 1 , \ldots , n
\end{equation}
In (\ref{jmp4cl}) we obviously have $\partial y_i /\partial x_j=b_{ij}$ for all $i,j=1, \ldots ,n$. Taking this into account, substitution of (\ref{jmp4mss}) in (\ref{jmp4jdiff}) implies that: 
\begin{equation}
\label{jmp4jjs}
      J^*_{ij}(y) = \sum_{p=1}^{r/2} \varphi_{2p-1}(y_{2p-1}) \varphi_{2p}(y_{2p}) 
	\sum_{k,l=1}^n b_{ik} b_{jl} \Lambda_{kl}^{2p-1,2p} 
\end{equation}
Using the definition (\ref{jmp4car}) in (\ref{jmp4jjs}) leads after some calculations to:
\begin{equation}
\label{jmp4jjs2}
      J^*_{ij}(y) = \sum_{p=1}^{r/2} \left| \begin{array}{cc} \delta_{i,2p-1} & \delta_{i,2p} 	\\ \delta_{j,2p-1} & \delta_{j,2p} \end{array} \right| 
	\varphi_{2p-1}(y_{2p-1}) \varphi_{2p}(y_{2p}) 
\end{equation}
In (\ref{jmp4jjs2}) three cases can be distinguished:
\begin{description}
\item[{\em Case 1.}] If it is $(i,j)=(2p-1,2p)$ we have $J^*_{ij}(y)=\varphi_i(y_i) 
\varphi_j(y_j)$. This is thus the case for $(i,j)=\{ (1,2), \ldots ,(r-1,r) \}$.
\item[{\em Case 2.}] When it is $(i,j)=(2p,2p-1)$ we find $J^*_{ij}(y)=- \varphi_i(y_i) \varphi_j(y_j)$. This happens for $(i,j)=\{ (2,1), \ldots ,(r,r-1) \}$.
\item[{\em Case 3.}] In any other situation, it is $J^*_{ij}(y)=0$.
\end{description}
Consequently, we have just arrived to the following structure matrix:
\begin{equation}
\label{jmp4jsy}
	{\cal J}^*(y)= 
	\left( \begin{array}{cc} 0 & \varphi_1 \varphi_2 \\
	- \varphi_1 \varphi_2 & 0 \end{array} \right) 
	\overbrace{ \oplus \ldots \oplus }^{r/2} 
	\left( \begin{array}{cc} 0 & \varphi_{r-1} \varphi_r  \\
	- \varphi_{r-1} \varphi_r & 0 \end{array} \right) 
	\oplus O_{1 \times 1} \overbrace{ \oplus \ldots \oplus }^{(n-r)} O_{1 \times 1}
\end{equation}
where $O_{1 \times 1}$ denotes the $1 \times 1$ null submatrix. Let us define the set 
$\Omega^* \subset I \! \! R^n$ according to $\Omega^* \equiv \{ B \cdot x : x \in \Omega \}$. It is clear that ${\cal J}^*(y)$ in (\ref{jmp4jsy}) is defined on $\Omega ^*$. Now let 
$y^*=(y_1^*, \ldots ,y_n^*) \in \Omega ^*$ be a point in which ${\cal J}^*(y)$ is evaluated. We then have $y^*=B \cdot x^*$ for some $x^* \in \Omega$. But this means that $y_i^*=B_i \cdot 
x^*$ for $i=1, \ldots ,r$, which implies that $y_i^* \in \Omega _i^*$ for all $i=1, \ldots ,r$. On the other hand, it is assumed by Definition 2.1 that every function $\varphi_i (y_i)$ does not vanish in $\Omega _i^*$ for $i=1, \ldots ,r$. We see then that Rank(${\cal J}^*$)$=r$ everywhere in $\Omega ^*$. Since according to transformation (\ref{jmp4jdiff}) matrices ${\cal J}(x)$ and ${\cal J}^*(y)$ are congruent, this implies in particular that Rank(${\cal J}$)$=r$ at every point of $\Omega$. Lemma 3.3 is thus demonstrated. \hfill Q.E.D.

\mbox{}

As a consequence of Lemmas 3.2 and 3.3, we have that the Casimir invariants (\ref{jmp4cas}) constitute a complete set. After this remark, the statements (a) and (b) of Theorem 3.1 are already demonstrated. Let us then regard item (c). The fact that Rank(${\cal J}$)$=r$ is constant in $\Omega$ implies that Darboux theorem is applicable. In the case $r=0$ the statement (c) of the theorem is valid since ${\cal J}$ does coincide with its Darboux canonical form, the diffeomorphic transformation thus being the identity. Then, in what remains of the proof we shall focus on the case $r \geq 2$. For this, the starting point will be matrix ${\cal J}^*(y)$ in (\ref{jmp4jsy}) which was obtained after the diffeomorphic transformation $y=B \cdot x$. Since every function $\varphi_i (y_i)$ does not vanish in $\Omega _i^*$ for $i=1, \ldots ,r$, 
it is possible to perform on ${\cal J}^*(y)$ an additional transformation of coordinates $z = z(y)$ defined as:
\begin{equation}
\label{jmp4tr2}
	\left\{ \begin{array}{lcl} 
	\displaystyle{z_i= \int \frac{\mbox{d}y_i}{\varphi_i(y_i)}} & , & i=1, \ldots ,r \\
	\mbox{} & \mbox{} & \mbox{} \\
	z_i = y_i & , & i=r+1, \ldots ,n
	\end{array} \right.
\end{equation}
Transformation (\ref{jmp4tr2}) is globally defined in $\Omega ^*$, and actually it is not difficult to verify that it is also diffeomorphic: since functions $\varphi_i(y_i)$ are $C^1$ and nonvanishing, both $z_i(y_i)$ and its inverse are always differentiable and strictly monotonic for every $i=1, \ldots ,n$. The outcome after transformation (\ref{jmp4tr2}) is a new structure matrix ${\cal J}^{**}(z)$ which is obtained from (\ref{jmp4jdiff}) and (\ref{jmp4tr2}) as:
\begin{equation}
\label{jmp4j2s}
      J^{**}_{ij}(z) = \sum_{k,l=1}^n \frac{\partial z_i}{\partial y_k} J^*_{kl}(y) 
	\frac{\partial z_j}{\partial y_l} = 
	\frac{\mbox{d} z_i}{\mbox{d} y_i} J^*_{ij}(y) \frac{\mbox{d} z_j}{\mbox{d} y_j}
	\:\; , \;\:\;\: i,j = 1 , \ldots , n
\end{equation}
Now two different cases are to be recognized:
\begin{description}
\item[{\em Case 1.}] If $1 \leq i \leq r$ and $1 \leq j \leq r$, then from (\ref{jmp4j2s}) we have:
\[
	J^{**}_{ij}(z)= \frac{J^*_{ij}(y)}{\varphi_i(y_i) \varphi_j(y_j)} \:\; , \:\;\:\; 
	i,j = 1, \ldots ,r
\]
\item[{\em Case 2.}] In any other case different to the previous one, we obtain $J^{**}_{ij}(z)=0$ because for all those values of $i$ and $j$ it is $J^*_{ij}(y)=0$ in expression 
(\ref{jmp4j2s}).
\end{description}
Accordingly, a comparison with (\ref{jmp4jsy}) shows that:
\begin{equation}
\label{jmp4jsz}
	{\cal J}^{**}(z)= 
	\left( \begin{array}{cc} 0 & 1 \\ -1 & 0 \end{array} \right) 
	\overbrace{ \oplus \ldots \oplus }^{r/2} 
	\left( \begin{array}{cc} 0 & 1 \\ -1 & 0 \end{array} \right) 
	\oplus O_{1 \times 1} \overbrace{ \oplus \ldots \oplus }^{(n-r)} O_{1 \times 1}
\end{equation}
Therefore the Darboux canonical form (\ref{jmp4jsz}) is globally constructed by means of a diffeomorphism for every $r \geq 2$. The proof of Theorem 3.1 is complete. \hfill Q.E.D.

\mbox{}

Thus the multiseparable Poisson structures considered, as well as their complete families of Casimir invariants and the global reduction to the Darboux canonical form, have been completely characterized after the previous results. At this stage, it is convenient to illustrate by means of some examples the generality of the family just analyzed as well as the different procedures described. This is the purpose of the next section.

\mbox{}

\begin{flushleft}
{\bf IV. EXAMPLES}
\end{flushleft}

We shall consider two different examples, well-known in the literature: the first one arises in population dynamics, while the second instance comes from mechanics.

\begin{flushleft}
{\bf Example 1. Kermack-McKendrick system}
\end{flushleft}

The following structure matrix is of interest$^{10,33\:}$ for the analysis of the 
Kermack-McKendrick model for epidemics: 
\begin{equation}
\label{jmp4kmk}
	{\cal J}(x) = Rx_1x_2 \left( \begin{array}{ccc} 0 & 1 & -1 \\ -1 & 0 & 1 \\ 1 & -1 & 0 
	\end{array} \right)
\end{equation}
where $R>0$ is a real constant. Since $x_i>0$ for all $i=1,2,3,$ it is Rank(${\cal J}$)$=2$, a Casimir invariant being $C(x)=x_1+x_2+x_3$. In terms of the elements described in Definition 2.1, matrix (\ref{jmp4kmk}) is multiseparable with:
\[
	A = \left( \begin{array}{ccc} 1 & 0 & 0 \\ 0 & 1 & 0 \\ -1 & -1 & 1 \end{array} \right) 
	\:\; , \;\:\: B = A^{-1} = 
	\left( \begin{array}{ccc} 1 & 0 & 0 \\ 0 & 1 & 0 \\ 1 & 1 & 1 \end{array} \right)	
\]
and functions $\varphi_i(y_i)= \kappa_i y_i$ for $i=1,2$, where $\kappa_1$ and $\kappa_2$ are arbitrary real constants verifying the condition $\kappa_1 \kappa_2=R$. We can check how ${\cal J}$ in (\ref{jmp4kmk}) is generated according to Definition 2.1:
\[
	J_{12}= \Lambda_{12}^{12} \varphi_1(B_1 \cdot x) \varphi_2(B_2 \cdot x)= 
	\left| \begin{array}{cc} 1 & 0 \\ 0 & 1 \end{array} \right| \kappa_1 x_1 \kappa_2 x_2 = 
	Rx_1x_2
\]
\[
	J_{13}= \Lambda_{13}^{12} \varphi_1(B_1 \cdot x) \varphi_2(B_2 \cdot x)= 
	\left| \begin{array}{cc} 1 & 0 \\ -1 & -1 \end{array} \right| \kappa_1 x_1 \kappa_2 x_2 = 
	-Rx_1x_2
\]
\[
	J_{23}= \Lambda_{23}^{12} \varphi_1(B_1 \cdot x) \varphi_2(B_2 \cdot x)= 
	\left| \begin{array}{cc} 0 & 1 \\ -1 & -1 \end{array} \right| \kappa_1 x_1 \kappa_2 x_2 = 
	Rx_1x_2
\]
The calculations for the remaining nonzero entries are entirely similar as far as 
$\Lambda_{ij}^{kl}=-\Lambda_{ji}^{kl}$ for all $i,j,k,l$. Let us now consider the Darboux canonical form for ${\cal J}$. If we apply (\ref{jmp4jdiff}) for the coordinate change 
(\ref{jmp4cl}), namely $y=B \cdot x$, we arrive after some calculations at: 
\begin{equation}
\label{jmp4kmc1}
	{\cal J}^*(y)= Ry_1y_2 \left( \begin{array}{ccc}
	0 & 1 & 0 \\ -1 & 0 & 0 \\ 0 & 0 & 0 \end{array} \right)
\end{equation}
To complete the reduction to the Darboux canonical form according to the procedure given in the previous section, an additional transformation (\ref{jmp4tr2}) is to be applied to matrix ${\cal J}^*(y)$ in (\ref{jmp4kmc1}). Now such transformation amounts to:
\begin{equation}
\label{jmp4kmc5}
	z_1 = \int \frac{\mbox{d}y_1}{\kappa_1 y_1} = \frac{1}{\kappa_1} \ln y_1 \;\: , \;\:\;\:
	z_2 = \int \frac{\mbox{d}y_2}{\kappa_2 y_2} = \frac{1}{\kappa_2} \ln y_2 \;\: , \;\:\;\:
	z_3 = y_3
\end{equation}
Then, the result after the change of coordinates (\ref{jmp4kmc5}) is the Darboux canonical form:
\[
	{\cal J}^{**}(z)= \left( \begin{array}{ccc}
	0 & 1 & 0 \\ -1 & 0 & 0 \\ 0 & 0 & 0 \end{array} \right)
\]
Consequently, the reduction is globally and constructively completed. The diffeomorphic character of all the transformations involved is also evident. 

\begin{flushleft}
{\bf Example 2. Poisson bracket for the Toda lattice}
\end{flushleft}

As a second example, a Poisson structure which is frequently employed for the study of the Toda system shall be considered$^{15\:}$. Toda lattice, when expressed in Flaschka's variables $x=(x_1, \ldots ,x_n)=(\alpha_1, \ldots , \alpha_{N-1}, \beta_1, \ldots , \beta_N)$ is a Poisson system with brackets 
\[
	\{ \alpha_i , \beta_i \} = - \alpha_i \:\; , \:\;\:\; 
	\{ \alpha_i , \beta_{i+1} \} = \alpha_i
\] 
while the rest of elementary brackets vanish. Therefore, this is a Poisson structure of dimension $n=2N-1$ and having the following structure matrix:
\begin{equation}
\label{jmp4tlf}
   {\cal J} = \left( \begin{array}{cccccccccc}
	\mbox{} & \mbox{} & \mbox{} & \mbox{} & \vline & 
		- \alpha_1 & \alpha_1 & \mbox{} & \mbox{} & \mbox{}  \\
	\mbox{} & \mbox{} & O_{(N-1) \times (N-1)} & \mbox{} & \vline & 
		\mbox{} & - \alpha_2 & \alpha_2 & \mbox{} & \mbox{} \\
	\mbox{} & \mbox{} & \mbox{} & \mbox{} & \vline & 
		\mbox{} & \mbox{} & \ddots & \ddots & \mbox{} \\
	\mbox{} & \mbox{} & \mbox{} & \mbox{} & \vline & 
		\mbox{} & \mbox{} & \mbox{} & - \alpha_{N-1} & \alpha_{N-1} \\ \hline
	\alpha_1 & \mbox{} & \mbox{} & \mbox{} & \vline & 
		\mbox{} & \mbox{} & \mbox{} & \mbox{} & \mbox{} \\
	- \alpha_1 & \alpha_2 & \mbox{} & \mbox{} & \vline & 
		\mbox{} & \mbox{} & \mbox{} & \mbox{} & \mbox{} \\
	\mbox{} & - \alpha_2 & \ddots & \mbox{} & \vline & 
		\mbox{} & \mbox{} & O_{N \times N} & \mbox{} & \mbox{} \\
	\mbox{} & \mbox{} & \ddots & \alpha_{N-1} & \vline & 
		\mbox{} & \mbox{} & \mbox{} & \mbox{} & \mbox{} \\
	\mbox{} & \mbox{} & \mbox{} & - \alpha_{N-1} & \vline & 
		\mbox{} & \mbox{} & \mbox{} & \mbox{} & \mbox{} 
      \end{array} \right)
\end{equation}
where $O$ denotes the null submatrix of size given by the subindex. It is immediate that the rank of ${\cal J}$ is $r=n-1=2N-2$. Consequently, there is only one independent Casimir invariant, namely $C(x) = \sum _{i=1}^{N} \beta_i$. 

Let us first show that the structure matrix (\ref{jmp4tlf}) is multiseparable for every $n \geq 3$. In terms of Definition 2.1, we now have the functions:
\begin{equation}
\label{jmp4etf}
	\left\{ \begin{array}{cclcl}
	\varphi_i(y_i) & = & -y_i & , & i=1,3, \ldots , r-1=n-2=2N-3 \\
	\varphi_i(y_i) & = &   1  & , & i=2,4, \ldots , r=n-1=2N-2 
	\end{array} \right.
\end{equation}
And the matrices $A$ and $B$ are given in what follows. In first place, we have for $A$:
\begin{equation}
\label{jmp4ta}
	A = \left( \begin{array}{ccccccccccc}
	-1 &  0 &  0 & 0 &  0 & 0 & \ldots &  0 & 0 & \vline & 0 \\
	 0 &  0 & -1 & 0 &  0 & 0 & \ldots &  0 & 0 & \vline & 0 \\
	 0 &  0 &  0 & 0 & -1 & 0 & \ldots &  0 & 0 & \vline & 0 \\
	\vdots & \vdots & \vdots & \vdots & \vdots & \vdots & 
		\mbox{} & \vdots & \vdots & \vline & \vdots \\
	 0 &  0 &  0 & 0 &  0 & 0 & \ldots & -1 & 0 & \vline & 0 \\ \hline
	 0 &  1 &  0 &  0 &  0 & 0 & \ldots &  0 & 0 & \vline & 0 \\
	 0 & -1 &  0 &  1 &  0 & 0 & \ldots &  0 & 0 & \vline & 0 \\
	 0 &  0 &  0 & -1 &  0 & 1 & \ldots &  0 & 0 & \vline & 0 \\
	\vdots & \vdots & \vdots & \vdots & \vdots & \vdots & 
		\mbox{} & \vdots & \vdots & \vline & \vdots \\
	 0 &  0 &  0 &  0 &  0 & 0 & \ldots &  0 & 1 & \vline & 0 \\
	 0 &  0 &  0 &  0 &  0 & 0 & \ldots &  0 & -1 & \vline & 1 
	\end{array} \right)
\end{equation}
Notice that for the sake of clarity, every row of $A$ is symbolically split in two parts of sizes $2N-2$ (left) and $1$ (right), while vertically every column is also divided schematically in two pieces of sizes $N-1$ (up) and $N$ (down). For $B$ we have:
\begin{equation}
\label{jmp4tb}
	B = \left( \begin{array}{ccccccccccccc}
	-1 &  0 &  0 & \ldots &  0 & \vline & 0 & 0 & 0 & 0 & \ldots & 0 & 0 \\
	 0 &  0 &  0 & \ldots &  0 & \vline & 1 & 0 & 0 & 0 & \ldots & 0 & 0 \\
	 0 & -1 &  0 & \ldots &  0 & \vline & 0 & 0 & 0 & 0 & \ldots & 0 & 0 \\
	 0 &  0 &  0 & \ldots &  0 & \vline & 1 & 1 & 0 & 0 & \ldots & 0 & 0 \\
	 0 &  0 & -1 & \ldots &  0 & \vline & 0 & 0 & 0 & 0 & \ldots & 0 & 0 \\
	 0 &  0 &  0 & \ldots &  0 & \vline & 1 & 1 & 1 & 0 & \ldots & 0 & 0 \\
	 \vdots & \vdots & \vdots & \mbox{} & \vdots & \vline & 
	 \vdots & \vdots & \vdots & \vdots & \mbox{} & \vdots & \vdots \\
	 0 &  0 & 0 & \ldots & -1 & \vline & 0 & 0 & 0 & 0 & \ldots & 0 & 0 \\
	 0 &  0 & 0 & \ldots &  0 & \vline & 1 & 1 & 1 & 1 & \ldots & 1 & 0 \\ \hline
	 0 &  0 & 0 & \ldots &  0 & \vline & 1 & 1 & 1 & 1 & \ldots & 1 & 1
	\end{array} \right)
\end{equation}
Again, for clarity every row of $B$ has been divided in two parts of sizes $N-1$ (left) and $N$ (right), while vertically every column is also separated in two pieces of sizes $2N-2$ (up) and $1$ (down). It is simple to check that $A$ in (\ref{jmp4ta}) and $B$ in (\ref{jmp4tb}) are invertible and $A = B^{-1}$. Let us verify that these elements generate the Poisson matrix (\ref{jmp4tlf}). According to Definition 2.1 and equations (\ref{jmp4etf}-\ref{jmp4tb}) we now have:
\begin{equation}
\label{jmp4etf2}
	\left\{ \begin{array}{cclcl}
	\varphi_i(B_i \cdot x) & = & \alpha_{(i+1)/2} & , & i=1,3, \ldots , r-1=n-2=2N-3 \\
	\varphi_i(B_i \cdot x) & = &   1  & , & i=2,4, \ldots , r=n-1=2N-2 
	\end{array} \right.
\end{equation}
Therefore using (\ref{jmp4mss}) together with (\ref{jmp4etf2}) we arrive at:
\begin{equation}
\label{jmp4tlc}
	J_{ij}(x)= \sum_{k=1}^{r/2} \Lambda_{ij}^{2k-1,2k} \alpha_{k}
	\:\; , \;\:\;\: i,j = 1 , \ldots , n
\end{equation}
If we examine matrix $A$ in (\ref{jmp4ta}) we see that four cases appear in (\ref{jmp4tlc}):
\begin{description}
\item[{\em Case 1:}] $1 \leq i \leq (N-1)$, $1 \leq j \leq (N-1)$. In this case, every determinant $\Lambda_{ij}^{2k-1,2k}$ contains at least three zeroes, and thus vanishes.
\item[{\em Case 2:}] $N \leq i \leq (2N-1)$, $N \leq j \leq (2N-1)$. Now every determinant $\Lambda_{ij}^{2k-1,2k}$ has a null column, and consequently also vanishes.
\item[{\em Case 3:}] $1 \leq i \leq (N-1)$, $N \leq j \leq (2N-1)$. Examination of $A$ shows that the coefficient $\Lambda_{ij}^{2k-1,2k}$ will be different from zero if and only if for a given $i$ it is $k=i$, and $j$ takes any of the two values $j=(i+N-1)$ or $j=(i+N)$. Then, according to (\ref{jmp4tlc}) the only entries of ${\cal J}$ that do not vanish are the ones associated to those determinants $\Lambda_{ij}^{2k-1,2k}$ that are not zero, which are:
\begin{equation}
\label{jmp4tll}
	\left\{ \begin{array}{lclcl}
	\Lambda_{i,i+N-1}^{2i-1,2i}=-1 & \Rightarrow & 
		J_{i,i+N-1}= - \alpha_i & , & i=1, \ldots , N-1 \\
	& & & & \\
	\Lambda_{i,i+N}^{2i-1,2i}= 1 & \Rightarrow & 
		J_{i,i+N}=  \alpha_i & , & i=1, \ldots , N-1 
	\end{array} \right.
\end{equation}
\item[{\em Case 4:}] $N \leq i \leq (2N-1)$, $1 \leq j \leq (N-1)$. This case is 
skew-symmetrical of Case 3, therefore it is not necessary to repeat the calculations since the argument is entirely similar.
\end{description}
The outcome of the previous classification is precisely matrix ${\cal J}$ in (\ref{jmp4tlf}), as expected. 

To conclude the example, let us now turn to the construction of the Darboux canonical form, developed in the last section. As we know, the first step is the coordinate transformation 
(\ref{jmp4cl}) of the form $y= B \cdot x$, where $y=(y_1, \ldots ,y_n)$. From the definition of $B$ in (\ref{jmp4tb}) note in particular that we now have:
\begin{equation}
\label{jmp4ytl}
	y_{2i-1} = - \alpha_i \:\; , \;\:\;\: i=1, \ldots ,N-1
\end{equation}
Making use of (\ref{jmp4tlc}), (\ref{jmp4tll}) and (\ref{jmp4ytl}), the application to ${\cal J}$ in (\ref{jmp4tlf}) of the transformation rule (\ref{jmp4jdiff}) for the change 
(\ref{jmp4cl}) leads after some algebra to:
\begin{equation}
\label{jmp4tmys}
	{\cal J}^*(y)= 
	\left( \begin{array}{cc} 0 & -y_1 \\ y_1 & 0 \end{array} \right) 
	\overbrace{ \oplus \ldots \oplus }^{(N-1)} 
	\left( \begin{array}{cc} 0 & -y_{2N-3} \\ y_{2N-3} & 0 \end{array} \right) 
	\oplus O_{1 \times 1}
\end{equation}
We now apply to (\ref{jmp4tmys}) the second transformation (\ref{jmp4tr2}) which now becomes:
\begin{equation}
\label{jmp4tltr2}
	\left\{ \begin{array}{lcl} 
	\displaystyle{z_i= - \int \frac{\mbox{d}y_i}{y_i}} = - \ln y_i & , 
		& i=1,3, \ldots , 2N-3 \\
	\mbox{} & \mbox{} & \mbox{} \\
	\displaystyle{z_i = \int \mbox{d}y_i = y_i} & , & i=2,4, \ldots ,2N-2 \\
	\mbox{} & \mbox{} & \mbox{} \\
	z_i = y_i & , & i=2N-1
	\end{array} \right.
\end{equation}
Taking (\ref{jmp4j2s}) into account, the application of transformation (\ref{jmp4tltr2}) to the structure matrix (\ref{jmp4tmys}) finally leads to the Darboux canonical form:
\[
	{\cal J}^{**}(z)= 
	\left( \begin{array}{cc} 0 & 1 \\ -1 & 0 \end{array} \right) 
	\overbrace{ \oplus \ldots \oplus }^{(N-1)} 
	\left( \begin{array}{cc} 0 & 1 \\ -1 & 0 \end{array} \right) 
	\oplus O_{1 \times 1} 
\]
Recall also how the diffeomorphic character of both coordinate transformations (\ref{jmp4cl}) and (\ref{jmp4tltr2}) is clear in practice.

\mbox{}

\begin{flushleft}
{\bf V. FINAL REMARKS}
\end{flushleft}

Investigation of skew-symmetric solutions of the Jacobi equations provides an increasingly rich perspective of finite-dimensional Poisson structures. In spite that a complete knowledge of such solutions is still far, the investigation of the problem seems to be not only a mathematically appealing subject, but also a unavoidable issue for a better understanding of finite-dimensional Poisson systems, and therefore of the scope of Hamiltonian dynamics. The validity of the previous statement holds from the fact that such knowledge provides a richer framework for the fundamental problem of recasting a given differential flow into a Poisson system, whenever possible, as well as an explicit link with classical Hamiltonian theory through the construction of the Darboux canonical form.

It is well-known that the skew-symmetric Jacobi equations become increasingly complex as dimension grows. This explains that the characterization of families of arbitrary dimension composed by generic functions (namely not limited to a given degree of nonlinearity),  having arbitrary rank and being amenable to a global and constructive analysis (including the determination of the Darboux canonical form) is still very uncommon in the literature. In addition, the characterization of such a new solution family often allows the conceptual and operational unification of diverse Poisson structures and systems previously known but unrelated, which can hereafter be regarded from a more general and economic standpoint. Examples of this have been given in the previous section. In particular, in such sense it is physically interesting to identify the Casimir invariants and to develop the reduction procedure to the Darboux canonical form for the new solution families. These are features of special relevance when they can be globally achieved, thus providing an additional instance of a result that goes beyond the {\em a priori \/} scope of Darboux theorem ---something reported only in a limited number of cases. This kind of results suggests that the direct investigation of the Jacobi equations constitutes a fruitful line of research not only for classification purposes but also for the detailed analysis of Poisson structures, not to mention its mathematical interest as an example of nonlinear system of PDEs. For these reasons, all of them just illustrated in the case of multiseparable solutions, this subject should deserve further attention in the future.

\pagebreak
\begin{flushleft}
{\bf References and notes}
\end{flushleft}
\begin{description}
\addtolength{\itemsep}{-0.2cm}
\item[$^1$] A. Lichnerowicz, J. Diff. Geom. {\bf 12}, 253 (1977); A. Weinstein, J. Diff. Geom. {\bf 18}, 523 (1983).  
\item[$^2$] P. J. Olver, {\em Applications of Lie Groups to Differential Equations}, 2nd ed. 
	(Springer-Verlag, New York, 1993).  
\item[$^{3}$] P. J. Morrison, Rev. Mod. Phys. {\bf 70}, 467 (1998).  
\item[$^{4}$] R. D. Hazeltine, D. D. Holm and P. J. Morrison, J. Plasma Phys. {\bf 34}, 103 
	(1985); D. D. Holm, Phys. Lett. A {\bf 114}, 137 (1986); P. J. Morrison and J. M. Greene, 
	Phys. Rev. Lett. {\bf 45}, 790 (1980). 
\item[$^{5}$] J. E. Marsden, R. Montgomery, P. J. Morrison and W. B. Thompson, Ann. Phys. 	(N.Y.) {\bf 169}, 29 (1986).
\item[$^{6}$] I. E. Dzyaloshinskii and G. E. Volovick, Ann. Phys. (N.Y.) {\bf 125}, 67 
	(1980).  
\item[$^7$] L. Cair\'{o} and M. R. Feix, J. Phys. A {\bf 25}, L1287 (1992). 
\item[$^{8}$] B. Hern\'{a}ndez--Bermejo and V. Fair\'{e}n, J. Math. Phys. {\bf 39}, 6162 
	(1998); B. Hern\'{a}ndez--Bermejo and V. Fair\'{e}n, J. Math. Anal. Appl. {\bf 256}, 242 
	(2001).
\item[$^{9}$] Y. Nutku, Phys. Lett. A {\bf 145}, 27 (1990).  
\item[$^{10}$] Y. Nutku, J. Phys. A {\bf 23}, L1145 (1990).  
\item[$^{11}$] M. Plank, J. Math. Phys. {\bf 36}, 3520 (1995); M. Plank, SIAM (Soc. Ind. Appl. 	Math.) J. Appl. Math. {\bf 59}, 1540 (1999).
\item[$^{12}$] M. Plank, Nonlinearity {\bf 9}, 887 (1996). 
\item[$^{13}$] F. Haas, J. Phys. A: Math. Gen. {\bf 35}, 2925 (2002).
\item[$^{14}$] K. Marciniak and S. Rauch-Wojciechowski, J. Math. Phys. {\bf 39}, 5292 (1998). 
\item[$^{15}$] P. A. Damianou, J. Math. Phys. {\bf 35}, 5511 (1994); 
		   P. A. Damianou, Rep. Math. Phys. {\bf 40}, 443 (1997);
		   P. A. Damianou, J. Geom. Phys. {\bf 45}, 184 (2003);
		   P. A. Damianou and S. P. Kouzaris, J. Phys. A: Math. Gen. {\bf 36}, 1385 (2003). 
\item[$^{16}$] S. A. Hojman, J. Phys. A {\bf 24}, L249 (1991); S. A. Hojman, J. Phys. A {\bf 	29}, 667 (1996); C. A. Lucey and E. T. Newman, J. Math. Phys. {\bf 29}, 2430 (1988); V. 	Perlick, J. Math. Phys. {\bf 33}, 599 (1992). 
\item[$^{17}$] R. G. Littlejohn, J. Math. Phys. {\bf 20}, 2445 (1979); R. G. Littlejohn, J. 	Math. Phys. {\bf 23}, 742 (1982); J. R. Cary and R. G. Littlejohn, Ann. Phys. (N.Y.) {\bf 	151}, 1 (1983).
\item[$^{18}$] D. David, D. D. Holm and M. V. Tratnik, Phys. Rep. {\bf 187}, 281 (1990).
\item[$^{19}$] G. Picard and T. W. Johnston, Phys. Rev. Lett. {\bf 48}, 1610 (1982). 
\item[$^{20}$] J. Goedert, F. Haas, D. Hua, M. R. Feix and L. Cair\'{o}, J. Phys. A {\bf 27}, 	6495 (1994). 
\item[$^{21}$] F. Haas and J. Goedert, Phys. Lett. A {\bf 199}, 173 (1995); 
	B. Hern\'{a}ndez--Bermejo and V. Fair\'{e}n, Phys. Lett. A {\bf 234}, 35 (1997).  
\item[$^{22}$] G. B. Byrnes, F. A. Haggar and G. R. W. Quispel, Physica A {\bf 272}, 99 
	(1999).
\item[$^{23}$] R. I. McLachlan, Phys. Rev. Lett. {\bf 71}, 3043 (1993);
	R. I. McLachlan, G. R. W. Quispel and N. Robidoux, Phys. Rev. Lett. {\bf 81}, 2399 	(1998).
\item[$^{24}$] D. D. Holm, J. E. Marsden, T. Ratiu and A. Weinstein, Phys. Rep. {\bf 123}, 1 	(1985). 
\item[$^{25}$] J. C. Simo, T. A. Posbergh and J. E. Marsden, Phys. Rep. {\bf 193}, 279 (1990). 
\item[$^{26}$] B. Hern\'{a}ndez--Bermejo and V. Fair\'{e}n, Phys. Lett. A {\bf 241}, 148 	(1998); T. W. Yudichak, B. Hern\'{a}ndez--Bermejo and P. J. Morrison, Phys. Lett. A {\bf 	260}, 475 (1999). 
\item[$^{27}$] D. David and D. D. Holm, J. Nonlinear Sci. {\bf 2}, 241 (1992). 
\item[$^{28}$] P. J. Olver, Phys. Lett. A {\bf 148}, 177 (1990); P. Gao, Phys. Lett. A {\bf 	273}, 85 (2000); C. Gonera and Y. Nutku, Phys. Lett. A {\bf 285}, 301 (2001).
\item[$^{29}$] R. G. Littlejohn, J. Plasma Phys. {\bf 29}, 111 (1983); P. Crehan, Prog. 	Theor. Phys. Suppl. {\bf 110}, 321 (1992).
\item[$^{30}$] K. Ngan, S. Meacham and P. J. Morrison, Phys. Fluids {\bf 8}, 896 (1996).
\item[$^{31}$] B. Hern\'{a}ndez--Bermejo, J. Math. Phys. {\bf 42}, 4984 (2001); B. 
	Hern\'{a}ndez--Bermejo, Phys. Lett. A {\bf 355}, 98 (2006); B. Hern\'{a}ndez--Bermejo, J. 	Math. Phys. {\bf 47}, 022901 (2006); B. Hern\'{a}ndez--Bermejo and V. Fair\'{e}n, Phys. 	Lett. A {\bf 271}, 258 (2000). 
\item[$^{32}$] K. H. Bhaskara and K. Rama, J. Math. Phys. {\bf 32}, 2319 (1991); 
	B. Hern\'{a}ndez--Bermejo, Phys. Lett. A {\bf 287}, 371 (2001);
	S. Lie, {\em Theorie der Transformationsgruppen\/} (B. G. Teubner, Leipzig, 1888); 
	Z.-J. Liu and P. Xu, Lett. Math. Phys. {\bf 26}, 33 (1992). 
\item[$^{33}$] H. G\"{u}mral and Y. Nutku, J. Math. Phys. {\bf 34}, 5691 (1993). 
\item[$^{34}$] A. Ay, M. G\"{u}rses and K. Zheltukhin, J. Math. Phys. {\bf 44}, 5688 (2003). 
\end{description}
\end{document}